\documentclass[journal]{IEEEtran}

\usepackage{cite}

\ifCLASSINFOpdf
  \usepackage[pdftex]{graphicx}
\else
\fi

\usepackage{acronym}
\usepackage{amsmath}
\allowdisplaybreaks
\usepackage{array}
\usepackage{float}
\usepackage{dotlessi}
\usepackage{bm}
\usepackage{xcolor}

\newacro{gfm}[GFM]{Grid-Forming}
\newacro{gfl}[GFL]{Grid-Following}
\newacro{tsc}[TSC]{Transient Slack Capability}
\newacro{ibr}[IBR]{Inverter-Based Resource}
\newacro{pll}[PLL]{Phase-Locked Loop}
\newacro{vsm}[VSM]{Virtual Synchronous Machine}
\newacro{sm}[SM]{Synchronous Machine}
\newacro{ph}[PH]{port-Hamiltonian}
\newacro{cf}[CF]{Complex Frequency}

\newcommand{\bfg}{\boldsymbol}
\renewcommand{\bm}{\mathbf}

\begin{document}

\title{Transient Slack Capability}

\author{%
  Rodrigo~Bernal,~\IEEEmembership{IEEE Member,} and
  Federico~Milano,~\IEEEmembership{IEEE Fellow}%
  \thanks{R.~Bernal and F.~Milano are with the School of Electrical and Electronic 
  Engineering, University College Dublin, Belfield Campus, D04V1W8, Ireland. e-mails: rodrigo.bernal@ucdconnect.ie, federico.milano@ucd.ie}%
  \thanks{This work is supported by Sustainable Energy Authority of Ireland (SEAI) by funding R.~Bernal and F.~Milano under project FRESLIPS, Grant No.~RDD/00681.}%
  \vspace{-7mm}
}

\maketitle

\begin{abstract}
    This paper introduces the concept of \ac{tsc}, a set of three necessary device-level conditions to ensure stability under sustained power perturbations. \ac{tsc} states that a device must (1) possess sufficient stored energy; (2) a controlled input power; and (3) maintain internal energy balance and synchronization. The paper shows that the relation among the time-scales of storage, control, and power perturbation is at the core of the \ac{tsc} concept.  Using the \ac{ph} framework, these conditions are formalized and validated via simulations on an adapted model of the WSCC 9-bus system. Case studies demonstrate that \ac{tsc} is achievable in both \ac{gfl} and \ac{gfm} converter control schemes, provided the conditions above are satisfied. Sensitivity analysis serves to identify storage and power reserve requirements to meet Conditions 1 and 2; the impact of converter current limiters on Condition 3; and inertia-less solutions able to achieve \ac{tsc}.  
\end{abstract}


\IEEEpeerreviewmaketitle

\section{Introduction}
\label{sec:intro}

Which specific dynamic feature of a generator is decisive in preventing system collapse under disturbances? Is it inertia, synchronization mechanism, a combination of both, or some entirely different property?  
To address these questions, we utilize an energy-based modeling framework to define the necessary conditions that a device has to satisfy to withstand a sustained power perturbation and drive the system to a new steady state.  We call this feature \textit{\acf{tsc}} and we show that if no device or set of devices in the system provides \ac{tsc}, the system cannot withstand a sustained power perturbation and will effectively collapse.

A device's ability to absorb, provide or redistribute energy during transients is essential for transient stability, this role it is typically linked to the slack bus concept.  In \cite{Dhople2020_slackbus} a distributed slack approach is proposed, demonstrating lower power flow mismatch errors than single slack methods.  In \cite{Milano2024_slackbus}, the author introduces a dynamic slack model with energy sources, identifying transient energy components and framing synchronous machines as distributed slack sources.  These works indicate a growing need to understand the transient energy exchange capabilities at the device level, especially as \acp{ibr} become dominant.

\acp{ibr} are typically categorized as either \acf{gfl} or \acf{gfm}, based on their control strategies.  \ac{gfl} inverters synchronize using \acp{pll} and regulate output current, making them dominant in today’s renewable energy integration. However, their reliance on external voltage references renders them susceptible to instability in weak grids \cite{Dong2015_PLL}. In \cite{Zhou2023_100} it is showed that small-signal stability of \ac{pll}-based inverters fundamentally depends on outer-loop damping and synchronizing torque. Despite these challenges, \ac{gfl} inverters can support even 100\% inverter-based power systems, with carefully tuned parameters \cite{Deepak2020_PLL100}. In contrast, \ac{gfm} inverters regulate local voltage and frequency, inherently providing inertial and damping characteristics that enhance stability in low-inertia conditions \cite{Matevosyan2019_GFM}.  Alternatives such as power synchronization control (PSC) \cite{Wu2019_PSC} aim to preserve \ac{gfm}-like behavior while improving transient performance in weak grids. In this context, \cite{Debry2019} defines \ac{gfm} behavior in terms of frequency smoothing and limiting RoCoF, both of which are important to avoid collapse under perturbations.

Literature also recognizes the limitations of strict \ac{gfl} or \ac{gfm} classifications. For instance, \cite{Li2022_Duality} illustrates a duality between \ac{gfl} and \ac{gfm} in terms of synchronization mechanisms (\ac{pll} vs. droop), control structure (current-forming vs.~voltage-forming), and associated stability trade-offs.  Controllers that unify these paradigms are gaining attention, such as the unified \ac{gfl}/\ac{gfm} controller proposed by \cite{Sijia2022_Unified}, which combines \ac{pll}-based synchronization with power-frequency droop control to transition between modes. Similarly, \cite{Bao2025_GFMGFLControl} introduces a fusion-\ac{gfm} strategy using \ac{vsm} models to merge \ac{gfl} and \ac{gfm} capabilities.  As noted in \cite{foundations, Wang2020_Stability}, the increasing diversity and complexity of \ac{ibr} controls highlight the urgent need for unified, technology-agnostic frameworks for assessing system stability.

\ac{ph} theory has emerged as a systematic framework to model power networks in an energy-based approach. By capturing storage, control, dissipation, and interconnection structure explicitly, \ac{ph} models offer a unified view of multi-physics power networks \cite{Fiaz2013}.  Most \ac{ph}-based research focuses on controller design and only a relatively small number of these works deals with modeling and stability issues \cite{Tonso2023_Survay}.  For instance, studies use \ac{ph} models to design droop or passivity controllers and to analyze specific networks \cite{Kong2024}.  In terms of stability analysis, a recent study proved global convergence of a \ac{ph}-modeled electromagnetic system under certain assumptions \cite{Xinyuan2025}.

In this work, we utilize the \ac{ph} framework to define device-level requirements for transient stability across different technologies and control strategies by introducing the concept of \acf{tsc}, a general energy-based set of conditions that determines whether a device can support system stability under disturbances.  \ac{tsc} enables technology-agnostic stability assessment and opens new opportunities for control design in \ac{ibr}-dominated systems.

This work provides the following contributions:
\begin{itemize}
\item \textit{Definition of \ac{tsc}}: we introduce and formalize the concept of \ac{tsc} as a set of three necessary conditions that a device must satisfy to support transient stability. If no device or set of devices in the system meet these conditions, the system cannot withstand sustained power perturbations and will effectively collapse. 
\item \textit{Interrelation of time-scales}: we show the inherent relation among the time-scales of storage, control, and power perturbations. We demonstrate that this interrelation is a fundamental aspect of the \ac{tsc} concept. 
\item \textit{Theoretical and numerical validation of \ac{tsc}}: we validate the \ac{tsc} framework through analytical assessment of \ac{gfl} and \ac{gfm} converter controllers as well as a comprehensive case study.
\end{itemize}
    
The remainder of the paper is organized as follows.  Section \ref{sec_Framework} introduces the framework necessary to understand the \ac{tsc}, which is then defined in detail in Section \ref{sec_TSCdef}.  Section \ref{sec_TSCassess} provides detailed examples of \ac{tsc} assessment for \ac{gfl} and \ac{gfm} models, along with insights into \acp{sm} and passive devices.  In Section \ref{sec_case}, six case studies are presented, including  three \ac{gfl} and three \ac{gfm} models simulated on a modified version of the well-known IEEE 9-bus system.  Finally, Section \ref{sec_conclusion} summarizes the conclusions and outlines future work.  
\vspace{-2mm}
\section{Framework}
\label{sec_Framework}

\subsection{\ac{ph} Structure of a Power System Device}

Power systems are commonly modeled in state space using a set of differential-algebraic equations (DAEs):
\begin{equation}
\begin{aligned}
   \label{eq:dae}
    \bm{x}' &= \bm{f}(\bm{x}, \bm{y}, t) \, , \\
    0 &= \bm{g}(\bm{x}, \bm{y}, t) \, .
\end{aligned}
\end{equation}
Here, $\bm{f}$ represents the differential equations that govern the time evolution of the full set of states $\bm{x}$, and $\bm{g}$ represents constraints involving the algebraic variables $\bm{y}$.  $\bm{x}'$ denotes the rate of change of the system states, and the variables $\bm{x}'$ and $\bm{y}$ become the degrees of freedom to be solved. For simplicity, time dependence $t$ is omitted in the remainder of the paper.

To apply \ac{ph} system theory to a specific device $k$, we define a storage function $\mathcal{H}_k(\bfg{x}_k)$ that measures the energy stored in a subset of the system states, $\bfg{x}_k \subset \bm{x}$. The total variation of this storage function is given by:
\begin{align}
    \mathcal{H}'_k(\bfg{x}_k) = \frac{\partial \mathcal{H}_k(\bfg{x}_k)}{\partial t} + \nabla \mathcal{H}_k(\bfg{x}_k)^{\top}  \bfg{x}'_k, \label{eq_dH}
\end{align}
where the partial derivative with respect to time captures any explicit time dependence of the storage function.

In this work, $\mathcal{H}_k(\bfg{x}_k)$ is interpreted as an absolute energy level, referenced to a nominal position of the states.  As a result, $\mathcal{H}'_k(\bfg{x}_k)$ reflects changes in the stored energy of the device, typically relative to an equilibrium state where $\mathcal{H}'_k(\bfg{x}_k) = 0$. For simplicity, the dependence on $\bfg{x}_k$ is omitted onward.

The dynamics of the device can then be expressed in the \ac{ph} form, as follows:
\begin{align}
    \bfg{x}'_k &= \left[\bfg{J}_k - \bfg{R}_k\right] \nabla \mathcal{H}_k + \bfg{G}_k \, \bfg{u}_k \label{eq_portH} \\
    \bfg{y}_k &= \bfg{G}_k^{\top} \nabla \mathcal{H}_k,
\end{align}
where $\bfg{J}_k$ is a skew-symmetric matrix representing internal energy-conserving interconnections, $\bfg{R}_k$ is a positive semi-definite matrix capturing internal dissipative phenomena, $\bfg{G}_k$ is the input-output matrix that maps control inputs $\bfg{u}_k$ to the state dynamics, and $\bfg{y}_k$ is the \ac{ph} output vector.

The matrix $\bfg{G}_k$ and the vector $\bfg{u}_k$ can be decomposed into components corresponding to different types of interactions:
\begin{align*}
    \bfg{G}_k = \begin{bmatrix}
        \bfg{G}^S_k & \bfg{G}^C_k & \bfg{G}^I_k
    \end{bmatrix}, \quad
    \bfg{u}_k^\top = \begin{bmatrix}
        \bfg{u}^S_k & \bfg{u}^C_k & \bfg{u}^I_k
    \end{bmatrix},
\end{align*}
where the superscripts $S$, $C$, and $I$ denote input source, control, and interconnection components, respectively. 

Throughout the remainder of this paper, we explore the physical and mathematical implications of \ac{ph} terms, analyzing their contributions to energy dynamics and their role in the system behavior.
\vspace{-3mm}
\subsection{Power Interface and Dirac Structures}
In the \ac{ph} framework, energy exchange is characterized by \emph{ports}, represented by pairs of \emph{flow} and \emph{effort} variables. For each device $k$, we define five types of ports: storage ports $(\bfg f_{\bfg{x}}, \bfg e_{\bfg{x}})$ with $\bfg f_{\bfg{x}} = -\bfg{x}'_k$ and $\bfg e_{\bfg{x}} = \nabla\mathcal{H}_k$, associated with internal energy storage; input ports $(\bfg f_{\bfg{u}^S}, \bfg e_{\bfg{u}^S})$ with $\bfg f_{\bfg{u}^S} = \bfg{u}^S_k$ and $\bfg e_{\bfg{u}^S} = \bfg{G}_k^{S\top}\nabla\mathcal{H}_k$, denoting external sources; control ports $(\bfg f_{\bfg{u}^C}, \bfg e_{\bfg{u}^C})$ with $\bfg f_{\bfg{u}^C} = \bfg{u}^C_k$ and $\bfg e_{\bfg{u}^C} = \bfg{G}_k^{C\top}\nabla\mathcal{H}_k$; interconnection ports $(\bfg f_{\bfg{u}^I}, \bfg e_{\bfg{u}^I})$ with $\bfg f_{\bfg{u}^I} = \bfg{u}^I_k$ and $\bfg e_{\bfg{u}^I} = \bfg{G}_k^{I\top}\nabla\mathcal{H}_k$, modeling the interconnection with the grid; and dissipation ports $(\bfg f_{\bfg{R}}, \bfg e_{\bfg{R}})$ with $\bfg f_{\bfg{R}} = \bfg{R}_k\nabla\mathcal{H}_k$ and $\bfg e_{\bfg{R}} = \nabla\mathcal{H}_k$, representing energy losses.

The power interface is formally described by Dirac structures, defined as subspaces $\mathcal{D} \subset \mathcal{F} \times \mathcal{E}$, where $\mathcal{F}$ denotes the space of {\it flows} and $\mathcal{E}$ the space of {\it efforts}. These structures ensure that all power exchanges, through storage, control, dissipation, and interconnection, remain balanced within the geometric framework. By taking all ports into account the power-balance extends to:
\begin{align}
     \bfg e_{\bfg{x}}^{\top}\bfg f_{\bfg{x}}+ \bfg e_{\bfg{u}^S}^{\top}\bfg f_{\bfg{u}^S}+ \bfg e_{\bfg{u}^C}^{\top}\bfg f_{\bfg{u}^C}+ \bfg e_{\bfg{u}^I}^{\top}\bfg f_{\bfg{u}^I}+ \bfg e_{\bfg{R}}^{\top}\bfg f_{\bfg{R}}=0.
\end{align}
\vspace{-8mm}

\subsection{Energy Balance as Equilibrium}

We define the energy balance in a system when it reaches a steady-state condition for its stored energy as follows:
\begin{align}
\mathcal{H}_{\mathrm{Sys}}' = \sum_{k \in \mathcal{K}} \mathcal{H}'_k = 0,
\end{align}
where $\mathcal{H}'_k = 0, \quad \forall k \in \mathcal{K}.$
Here, $\mathcal{H}_{\mathrm{Sys}}$ denotes the total internal stored energy of all the devices in the system, and $\mathcal{K}$ represents the set of devices $k$ that compose the system.

If a device $k$ does not have energy storage or its storage is either negligible (e.g., a transducer) or sufficiently large (e.g., a constant source) such that it can be considered on a different time scale, then its stored energy function and its time derivative are assumed to be null.

Since this work focuses on a device level perspective, from this point onward, $\mathcal{H}$ will explicitly denote the storage function of the device under study, omitting the subscript $k$.

\subsection{Conditions to Synchronize through Interconnection Port}

The sinusoidal parametrization and the resulting complex representation of power provide a comprehensive framework for understanding power flows in ac systems. This approach also highlights how energy storage elements contribute to alternating power delivery and absorption, an essential feature of reactive power.

Assuming a balanced three-phase system, steady-state conditions applied to the complex power injection at every bus, namely $\bar{s} = \bar{v} \bar{\imath}^*$, leads to:
\begin{align}
    \overline{s}'=&\   \frac{d}{dt} \left(v \imath\cos(\theta_{v}-\theta_{\imath})\right)+ \jmath \frac{d}{dt}\left( v \imath\sin(\theta_{v}-\theta_{\imath}) \right) = 0, 
\end{align}
where $\jmath$ denotes the imaginary unit. As each component has to be zero, it follows that:
\begin{align}
    \frac{d}{dt} \left(v \imath\right) = v' \imath+v \imath'= 0 &\quad\text{and}\quad
    \frac{d}{dt} \left(\theta_{v}-\theta_{\imath}\right) = 0.
\end{align}
If $v$ and $\imath$ are not null and do not tend to 0 or infinity, then:
\begin{align}
     v'/v = -\imath'/\imath .
\end{align}
which, at equilibrium, leads to:
\begin{align}
   \label{eq:vdotidot}
     v'\xrightarrow{} 0, \quad \imath'\xrightarrow{} 0.
\end{align}
Defining $\rho_v = v'/v$ and $\rho_{\imath} = \imath'/\imath$, then \eqref{eq:vdotidot} is equivalent to:
\begin{align}
   \label{eq:vdotidot2}
     \rho_v \xrightarrow{} 0, \quad \rho_{\imath} \xrightarrow{} 0.
\end{align}
For the angles, we have the following condition:
\begin{align}
    \theta_{v}'-\theta_{\imath}'=
    \omega_{v}-\omega_{\imath}\xrightarrow{} 0. 
\end{align}
If neither $\omega_{v}$ nor $\omega_{\imath}$ tend to zero or infinity, one has:
\begin{align}
    \omega_{v}=\omega_{\imath}=\omega_{s},
\end{align}
where $\omega_{s}$ is the synchronous frequency, which can be variable, but in practice a constant value is used. These conditions for synchronism are consistent with the ones proposed in \cite{Ponce2025}.   These conditions allow the angles to drift away, as long as its difference tends to zero. On the contrary, the magnitudes of the voltage and current are bounded to reach a constant value.

The quantities $\rho_v$, $\rho_{\imath}$, $\omega_v$ and $\omega_{\imath}$ are the real and imaginary parts of the complex frequency of the voltage and currents, respectively, as defined in \cite{cmplx}.  We use these quantities in Section \ref{sec_TSCassess} as they are convenient for analytical appraisal of the \ac{tsc} of \ac{gfl} and \ac{gfm} converters. 

\subsection{Time Scales}

By including the time derivative of a state in \eqref{eq_dH}, we inherently imply a time coupling between $\mathcal{H}'$ and the state variables. This means that changes in the measured energy $\mathcal{H}$ are linked to the rate of change of certain states, indicating that these states evolve on a similar time scale.
    
However, if the term involving the time derivative of a particular state is zero (i.e., $x'_i = 0$), the state is said to be time-decoupled. In this case, the state operates on a time scale that is entirely different from other dynamic processes in the system. This decoupling suggests that the state evolves either much faster or much slower than the processes.

    Faster state dynamics are often associated with low energy storage capacity within the states. As a result, the state acts as a transducer of other dynamics, effectively acting as an algebraic variable. On the other hand, slower dynamics are typically linked to a large energy storage capacity,     where the energy storage decouples its dynamics from the other states, such as in the case of a constant source.
    
\section{\ac{tsc} definition}

\label{sec_TSCdef}

The \ac{tsc} is defined as a set of three model-based conditions for a device connected to the grid that enables time continuous power balance under sustained power perturbations, consistently driving the system to a steady state energy balance. 

\subsection{Definition of Sustained Power Perturbation}

In the following subsections, we describe and define the three conditions.  However before getting into the theoretical developments, it is important to define the concept of \textit{sustained power perturbation}, which refers to a step change in grid power exchange that remains constant in the time scale of storage and control dynamics of the considered device.  If the power perturbation is not sustained, satisfying the three conditions discussed below does not guarantee \ac{tsc}.

\subsection{Condition 1: Storage Capacity}

Energy conversion in power systems is closely related to energy storage and its dynamics. For example, when the kinetic energy of a river’s flow is converted into the kinetic energy of a rotor, a dynamic relation is established.  This relation governs energy conversion through the coupled dynamics of both systems. The rotor’s motion depends on its resistance to changes in rotational speed, a property called inertia. Inertia allows the rotor to store rotational energy temporarily, enabling conversion of the water flow’s energy.

Consider a scenario where the rotor's inertia is very small. In this case, the rotor's motion would closely follow the dynamics of the water, resulting in negligible mechanical energy being stored in the rotor. The rotor's angular speed would simply mirror the energy stored in the water's motion. Conversely, if the rotor's inertia were infinitely large, the rotor's movement would remain virtually unaffected by the water flow. Thus, the rotor could theoretically store an infinite amount of energy, effectively decoupling the water dynamics from any system connected to the rotor.

These two extreme cases, no inertia and infinitely large inertia, represent the limiting conditions for energy storage. The first acts as a purely passive transducer, while the second acts as an idealized power slack. While neither of these extremes exist in reality, they are useful for modeling scenarios where the relative energy scales of storage devices differ significantly. When this occurs, the systems operate on different time scales.

Energy storage enables coupling across time scales and acts as a transient power buffer. Without storage capacity, a device cannot provide or absorb power on the same time scale as a power imbalance caused by a disturbance \cite{Milano2019_Storage}.

We define the storage capability of a device's model if it is possible to formulate an {energy storage function} $\mathcal{H}$, such that it satisfies the following properties:

\begin{enumerate}
    \item {Positive Semi-Definiteness}:
    \begin{align}
        \mathcal{H} \geq 0 \quad \forall \bfg{x}, \ \text{and} \quad \mathcal{H} = 0 \iff \bfg{x} = \bfg{0} \, .
    \end{align}
    This ensures that the stored energy is non-negative and zero only at the origin.

    \item {Smoothness}: The storage function $\mathcal{H}$ is of class $C^1$, ensuring the existence of its gradient and time derivative.

    \item {Passivity}: Characterizes systems whose internal energy never exceeds the energy supplied by their environment.
    \begin{align}
        {\mathcal{H}}'\leq \bfg{u}^{\top} \bfg{y},
    \end{align}
    where $\bfg{u}^{\top}\bfg{y}$ denotes the input and output ports of the device.  Furthermore, as shown in \cite{Schaft2014_PHoverview}, if a linear system with a quadratic storage function $\mathcal{H} = \frac{1}{2} \bfg{x}^\top \bfg Q \bfg{x}$ satisfies $\bfg Q \geq 0$, then it admits a \ac{ph} representation and is passive.
    \item {Radial Unboundedness}: This property allows a finite bound for all values of $\bfg{x}$.
    \begin{align}
        \mathcal{H} \to \infty \quad \text{as} \quad \|\bfg{x}\| \to \infty.
    \end{align}
    
\end{enumerate}

It is interesting to note that devices may therefore possess \ac{tsc} without relying on inertia as storage, as alternative forms of energy storage may satisfy the requirements of Condition 1 independently.

\subsection{Condition 2: Controlled Input Power}

A controlled input power is defined when an input variable $\bfg{u}^S$ influences the dynamics of the storage states of the device $\bfg{x}$.  In the \ac{ph} framework, this corresponds to the term $\bfg{G}^S\, \bfg{u}^S$.  The variable $\bfg{u}^S$ is not directly part of the stored energy function of the device, but it affects the system's dynamics through its influence on $\bfg{x}$.  Furthermore, $\bfg{u}^S$ can itself be a dynamic state, described by:
\begin{align}
    (\bfg{u}^{S})' = \bfg{f}_{{\bfg u}^{S}}\left(\bfg{x}, \bfg{u}\right). \label{eq_C2}
\end{align}

If $\bfg{G}^S\bfg{u}^S$ exists and the dynamics of $\bfg{u}^S$ is defined by (\ref{eq_C2}), then Condition 2 is satisfied.
Since $\bfg{u}^S$ behaves like a state, it remains unchanged at the instant the power perturbation occurs.  However, it contributes to the transient response and plays a crucial role in achieving a new steady-state energy balance by providing most of the required power compensation.

As mentioned in Condition 1, a constant input power can be seen as the limit case of an energy storage with an infinite amount of stored energy.  In this scenario, the input variable $\bfg{u}^S$ remains constant.  Even if it does not supply additional power, a steady-state energy balance can still be achieved, as long as there is enough dissipation in the device to provide power compensation.  In practice, dissipation is often minimized to improve efficiency, so although input variables may be slow, they are required to have a controlled dynamic.

Finally, note that satisfying Condition 2 is a necessary condition for a device to possess the \textit{dynamic slack bus capability} defined in \cite{Milano2024_slackbus}.

\subsection{Condition 3: Control-driven Energy Balance}

Conditions 1 and 2 do not explicitly address control objectives, while it is challenging to establish stability conditions without detailed knowledge of the system or the nature of the perturbation, we can define control objectives that directly or indirectly aim to drive the system toward an energy balance along feasible operating trajectories. 

A device is said to satisfy Condition 3 if its physical and control systems are designed to achieve a steady-state energy balance.  Note that Condition 3 does not imply stability.  It only refers to the ability of the controllers to drive the interface variables $\bfg e_I$ and $\bfg f_I$ toward a power balance condition (equivalent to power synchronization), while also maintaining internal energy balance. Therefore, the following conditions must be satisfied:
    \begin{align}
        \lim_{t\xrightarrow{}\infty}\mathcal{H}'=&0, \label{eq_C3_H}\\
        \lim_{t\xrightarrow{}\infty}\omega_v=&\lim_{t\xrightarrow{}\infty}\omega_{\imath}=\omega_s,\label{eq_C3_omega}\\
        \lim_{t\xrightarrow{}\infty}\rho_v=&\lim_{t\xrightarrow{}\infty}\rho_{\imath}=0, \label{eq_C3_rho}
    \end{align}
    where $\omega_s$ is bounded.

While Condition 2 characterizes input variables that do not influence the immediate response against a power perturbation, instantaneous power response is always required to maintain power balance. In this context, the only power available within the device (in addition to dissipation) is the stored energy defined by Condition 1. In this context, Condition 3 implicitly imposes the need of a power interface that acts as a power transducer, a system with negligible energy storage (as opposed to a power slack), to mediate power exchange between system perturbations, stored energy, dissipation, and inputs.  This power interface describes the architecture of the interaction for all physical and control interconnections.

The boundary conditions for the power interface are determined by the limits of the device's control and physical variables: $\bfg{x}^{\min} \leq \bfg{x} \leq \bfg{x}^{\max}$,    $\bfg{u}^{\min} \leq \bfg{u} \leq \bfg{u}^{\max}$ and $\bfg{y}^{\min} \leq \bfg{y} \leq \bfg{y}^{\max}$. Where \(\bfg{x}^{\min}\), \(\bfg{u}^{\min}\), \(\bfg{y}^{\min}\) and \(\bfg{x}^{\max}\), \(\bfg{u}^{\max}\), \(\bfg{y}^{\max}\) refer to their lower and upper bounds.

Consequently, if the device reaches any of these boundary conditions, it may no longer provide or absorb power, effectively decoupling its power response from the system and thereby losing its \ac{tsc} capability.

\subsection{Summary of Conditions}

These are the conditions that define if a device has \ac{tsc}:

\subsubsection{\bf Condition 1} ensures that there is sufficient stored energy within the device acting on the same time scale as the power perturbation.
\subsubsection{\bf Condition 2} ensures that there is access to stored energy outside the device acting in a time scale equal or greater than that of the power perturbation.
\subsubsection{\bf Condition 3} ensures internal energy balance and synchronization with the system. 

\ac{tsc} can then be interpreted in terms of time scales as the ability of a device to respond to a power imbalance by delivering or absorbing power on the same time scale as the perturbation that caused the imbalance while driving the system to a new energy balance. In other words, a device with \ac{tsc} can react quickly enough to counteract the disturbance within the time frame in which the imbalance is detected and access fast enough to other sources of stored energy to maintain synchronism and internal energy balance. 

\section{Examples of \ac{tsc} Model Assessment}
\label{sec_TSCassess}

This section presents examples of \ac{gfl} and \ac{gfm} models in detail and analyzes their \ac{tsc} capability by evaluating each of the three \ac{tsc} conditions.  Remarks on the 2nd order \ac{sm}, passive loads and storage devices are also provided. 

\subsection{Grid-following Converter}

We use the formulation presented in \cite{Diptak2023_GFL} to model and assess the \ac{gfl} converter, where the authors provide a detailed circuit representation along with its corresponding energy function.
Minor modifications are applied to the original model. In our analysis of the device, the grid is not taken into account.
\subsubsection{Model}
\begin{itemize}
    \item dc-bus voltage
    \begin{align}
        C_{\mathrm{dc}}v_{\mathrm{dc}}'=\imath_{\mathrm{dc}}^{\star}-\imath_{\mathrm{dc}}, \quad
        v_{\mathrm{dc}}\imath_{\mathrm{dc}}&= v_{\mathrm{d}}^t\imath_{\mathrm{d}}+v^t_{\mathrm{q}}\imath_{\mathrm{q}}. \label{eq_gfl_vdc}
    \end{align}
    
    \item Output Filter
    \begin{align}
        L_{\mathrm{f}} \imath_{\mathrm{d}}'&=-R_{\mathrm{f}} \imath_{\mathrm{d}}+\omega  L_{\mathrm{f}} \imath_{\mathrm{q}}+\left( v_{\mathrm{d}}^{\mathrm{t}}-v_{\mathrm{d}}\right),\label{eq_gfl_id}\\
        L_{\mathrm{f}} \imath_{\mathrm{q}}'&=-R_{\mathrm{f}} \imath_{\mathrm{q}}-\omega  L_{\mathrm{f}} \imath_{\mathrm{d}}+\left( v_{\mathrm{q}}^{\mathrm{t}}-v_{\mathrm{q}}\right)\label{eq_gfl_iq},\\
        C_{\mathrm{f}} v_{\mathrm{d}}'&=\omega C_{\mathrm{f}} v_{\mathrm{q}}+\left( \imath_{\mathrm{d}}-\imath_{\mathrm{d}}^{\mathrm{g}}\right),\label{eq_gfl_vd}\\
        C_{\mathrm{f}} v_{\mathrm{q}}'&=-\omega C_{\mathrm{f}} v_{\mathrm{d}}+\left( \imath_{\mathrm{q}}-\imath_{\mathrm{q}}^{\mathrm{g}}\right)\label{eq_gfl_vq}.
    \end{align}
    \item Power Interface
    \begin{align}
        v_{\mathrm{D}} = v_h \cos{\left(\theta_{\mathrm{g}}\right)}, \quad
        v_{\mathrm{Q}} = v_h \sin{\left(\theta_{\mathrm{g}}\right)},\\
        p_h = v_{\mathrm{d}}\imath_{\mathrm{d}}^{\mathrm{g}}+v_{\mathrm{q}}\imath_{\mathrm{q}}^{\mathrm{g}}, \quad
        q_h = v_{\mathrm{q}}\imath_{\mathrm{d}}^{\mathrm{g}}-v_{\mathrm{d}}\imath_{\mathrm{q}}^{\mathrm{g}}.\label{eq_gfl_pq}
    \end{align}
    \item dc-bus voltage controller
    \begin{align}
        \gamma_{\mathrm{DVC}}'&=v_{\mathrm{dc}}-v_{\mathrm{dc}}^{\star},\label{eq_gfl_gammaDVC} \\
        \imath_{\mathrm{d}}^{\star} &=K_{\mathrm{DVC}}^{\mathrm{p}}\left(v_{\mathrm{dc}}-v_{\mathrm{dc}}^{\star}\right)+K_{\mathrm{DVC}}^{\mathrm{i}}\gamma_{\mathrm{DVC}}.\label{eq_gfl_idref}
    \end{align}

    \item ac-bus voltage controller
    \begin{align}
        \gamma_{\mathrm{AVC}}'&=v_{\mathrm{ac}}-v_{\mathrm{ac}}^{\star},\\
        \imath_{\mathrm{q}}^{\star}&=K_{\mathrm{AVC}}^{\mathrm{p}}\left(v_{\mathrm{ac}}-v_{\mathrm{ac}}^{\star}\right)+K_{\mathrm{AVC}}^{\mathrm{i}}\gamma_{\mathrm{AVC}}.\label{eq_gfl_iqref}
    \end{align}
    \item Current controllers
    \begin{align}
        \gamma_{\mathrm{d}}'=&\imath_{\mathrm{d}}^{\star}-\imath_{\mathrm{d}}, \quad \gamma_{\mathrm{q}}'=\imath_{\mathrm{q}}^{\star}-\imath_{\mathrm{q}},\\
        v_{\mathrm{d}}^{\mathrm{t}}=&K_{\mathrm{CC}}^{\mathrm{p}}\left(\imath_{\mathrm{d}}^{\star}-\imath_{\mathrm{d}}\right)+K_{\mathrm{CC}}^{\mathrm{i}}\gamma_{\mathrm{d}} -\omega L_{\mathrm{f}} \imath_{\mathrm{q}}+v_{\mathrm{d}}, \label{eq_gfl_vtd}\\
        v_{\mathrm{q}}^{\mathrm{t}}=&K_{\mathrm{CC}}^{\mathrm{p}}\left(\imath_{\mathrm{q}}^{\star}-\imath_{\mathrm{q}}\right)+K_{\mathrm{CC}}^{\mathrm{i}}\gamma_{\mathrm{q}} +\omega L_{\mathrm{f}} \imath_{\mathrm{d}}+v_{\mathrm{q}}. \label{eq_gfl_vtq}
    \end{align}
    \item Phase-locked Loop
    \begin{align}
        \theta &= \theta_{\mathrm{e}}-\theta_g, \quad
        \theta_{\mathrm{e}}'=\omega, \quad \gamma_{\mathrm{PLL}}'= v_{\mathrm{q}},\label{eq_gfl_theta}\\
        \omega &= K_{\mathrm{PLL}}^{\mathrm{p}}v_{\mathrm{q}}+ K_{\mathrm{PLL}}^{\mathrm{i}}\gamma_{\mathrm{PLL}}+\omega_s.  \label{eq_gfl_omega}
    \end{align}
\end{itemize}

It is noteworthy that as the phase reference is external, a \ac{pll} model is required introducing a shift in the local voltages and currents with respect to the grid phase, specifically a shift denoted by $\theta$, as expressed in (\ref{eq_gfl_theta}). This shift maintains power invariance, meaning that any power required by the grid in system coordinates (DQ) is instantly transmitted to the local coordinates (dq).

\subsubsection{\ac{tsc} assessment}
We define the storage states as follows:
\begin{align}
    \bfg{x}=\begin{bmatrix}v_{\mathrm{dc}} & \imath_{\mathrm{d}}& \imath_{\mathrm{q}} & v_{\mathrm{d}}& v_{\mathrm{q}} \end{bmatrix}^{\top} \label{eq:gfl:x}
\end{align}

The storage function is then given by the capacitance at the dc side, and the filter inductance and capacitance at the ac side, as follows:
\begin{align}
   \mathcal{H}= \frac{C_{\mathrm{dc}} v_{\mathrm{dc}}^2}{2}+\frac{L_{\mathrm{f}}\left( \imath_{\mathrm{d}}^2+\imath_{\mathrm{q}}^2\right)}{2}+\frac{C_{\mathrm{f}}\left( v_{\mathrm{d}}^2+v_{\mathrm{q}}^2\right)}{2}.\label{eq:gfl:Hx}
\end{align}

This function is a quadratic function of the storage states.  Thus, $\mathcal{H}$ is a valid storage function, and therefore, {\bf Condition 1 is satisfied}. However, the total energy stored in the dc capacitor and ac filter is so small that is negligible in practice.  As a result, power balance is predominantly provided by the rapid response of the input current, as will be discussed in Section~\ref{sec_stored_vs_input}.

The time derivative of the storage function is given by:
\begin{align}
    \mathcal{H}'&=-{\bfg e}_S^{\top} {\bfg f}_S=-\left(\nabla\mathcal{H}\right)^\top {\bfg{x}'}\\
    =&-C_{\mathrm{dc}}v_{\mathrm{dc}}v_{\mathrm{dc}}'-L_{\mathrm{f}}\left(\imath_{\mathrm{d}}\imath_{\mathrm{d}}'-\imath_{\mathrm{q}}\imath_{\mathrm{q}}' \right)
     -C_{\mathrm{f}}\left(v_{\mathrm{d}}v_{\mathrm{d}}'-v_{\mathrm{q}}v_{\mathrm{q}}'\right) \notag.
\end{align}

By taking the gradient of (\ref{eq:gfl:Hx}) with respect to the storage states we obtain the following:
\begin{align}
    \nabla\mathcal{H}= \begin{bmatrix} C_{\mathrm{dc}}v_{\mathrm{dc}} &  L_{\mathrm{f}}\imath_{\mathrm{d}} &  L_{\mathrm{f}}\imath_{\mathrm{q}}&  C_{\mathrm{f}}v_{\mathrm{d}} &  C_{\mathrm{f}}v_{\mathrm{q}} \end{bmatrix} ^{\top}. \label{eq:gfl:gradH}
\end{align}

The input variables from the slack power controller is given by $
    \bfg{u}^S= 
\begin{bmatrix} 
    \imath^{\star}_{\mathrm{dc}}
\end{bmatrix}, \label{eq:gfl:uS}
$ and its dynamic for this model is constant.

All reference set points and control states can be represented through the  control variables $
    \bfg{u}^C= 
    \begin{bmatrix} 
        v^{t}_{\mathrm{d}} & v^{t}_{\mathrm{q}} & \omega
    \end{bmatrix}. \label{eq:gfl:uC}$
The interface power variables are given by the output current of the filter $
    \bfg{u}^I = 
    \begin{bmatrix} 
        \imath_{\mathrm{d}}^{\mathrm{g}} &  
        \imath_{\mathrm{q}}^{\mathrm{g}} 
    \end{bmatrix} ^{\top}.\label{eq:gfl:uI}$
By rearranging the storage state dynamics so we factorize by (\ref{eq:gfl:gradH}), we end up with:
\begin{align}
v_{\mathrm{dc}}'&= \begin{bmatrix} 0 &  \frac{-v^{t}_{\mathrm{d}} }{ L_{\mathrm{f}} C_{\mathrm{dc}} v_{\mathrm{dc}}}  & \frac{-v^{t}_{\mathrm{q}} }{ L_{\mathrm{f}} C_{\mathrm{dc}} v_{\mathrm{dc}}} & 0 & 0\end{bmatrix}
\nabla\mathcal{H}+ \frac{\imath^{\star}_{\mathrm{dc}}}{C_{\mathrm{dc}}},  \label{eq:gfl:Jvdc}\\
\imath_{\mathrm{d}}'&= \begin{bmatrix} \frac{v^{t}_{\mathrm{d}} }{ L_{\mathrm{f}} C_{\mathrm{dc}} v_{\mathrm{dc}}} &  -\frac{R_{\mathrm{f}}}{ L_{\mathrm{f}}^2 }  &\frac{\omega}{ L_{\mathrm{f}}} &-\frac{1}{L_{\mathrm{f}}C_{\mathrm{f}}} & 0\end{bmatrix}
\nabla\mathcal{H},\\
\imath_{\mathrm{q}}'&= \begin{bmatrix} \frac{v^{t}_{\mathrm{q}} }{ L_{\mathrm{f}} C_{\mathrm{dc}} v_{\mathrm{dc}}} &  -\frac{\omega}{ L_{\mathrm{f}}} &-\frac{R_{\mathrm{f}}}{ L_{\mathrm{f}}^2 } &0& -\frac{1}{L_{\mathrm{f}}C_{\mathrm{f}}} \end{bmatrix}
\nabla\mathcal{H},\\
v_{\mathrm{d}}'&= \begin{bmatrix} 0 & \frac{1}{L_{\mathrm{f}}C_{\mathrm{f}}}  &0 &0& \frac{\omega}{C_{\mathrm{f}}} \end{bmatrix}
\nabla\mathcal{H}-\frac{\imath^{\mathrm{g}}_{\mathrm{d}}}{C_{\mathrm{f}}},\\
v_{\mathrm{q}}'&= \begin{bmatrix} 0 & 0 & \frac{1}{L_{\mathrm{f}}C_{\mathrm{f}}}  &- \frac{\omega}{C_{\mathrm{f}}} & 0\end{bmatrix}
\nabla\mathcal{H}-\frac{\imath^{\mathrm{g}}_{\mathrm{q}}}{C_{\mathrm{f}}}. \label{eq:gfl:Jvq}
\end{align}

From the dynamic of the storage states (\ref{eq:gfl:Jvdc}) to (\ref{eq:gfl:Jvq}), we can form the matrices $\bfg{J}$, $\bfg{R}$, and $\bfg{G}$ as follows:
\begin{align}
\bfg{J}= &
\begin{bmatrix} 
    0 &  \frac{-v^{t}_{\mathrm{d}} }{ L_{\mathrm{f}} C_{\mathrm{dc}} v_{\mathrm{dc}}}  & \frac{-v^{t}_{\mathrm{q}} }{ L_{\mathrm{f}} C_{\mathrm{dc}} v_{\mathrm{dc}}} & 0 & 0\\
    \frac{v^{t}_{\mathrm{d}} }{ L_{\mathrm{f}} C_{\mathrm{dc}} v_{\mathrm{dc}}} &  0  &\frac{\omega}{ L_{\mathrm{f}}} &\frac{-1}{L_{\mathrm{f}}C_{\mathrm{f}}} & 0\\
    \frac{v^{t}_{\mathrm{q}} }{ L_{\mathrm{f}} C_{\mathrm{dc}} v_{\mathrm{dc}}} & \frac{-\omega}{ L_{\mathrm{f}}}& 0 &0& \frac{-1}{L_{\mathrm{f}}C_{\mathrm{f}}} \\
    0 & \frac{1}{L_{\mathrm{f}}C_{\mathrm{f}}}  &0 &0& \frac{\omega}{C_{\mathrm{f}}} \\
    0 & 0 & \frac{1}{L_{\mathrm{f}}C_{\mathrm{f}}}  & \frac{-\omega}{C_{\mathrm{f}}} & 0
\end{bmatrix}
\end{align}
\begin{align}
    \bfg{R} = &
    \begin{bmatrix}
    0 & 0 & 0 & 0 & 0\\
    0 & \frac{R_{\mathrm{f}}}{L_{\mathrm{f}}^2} & 0  & 0 & 0\\
    0 & 0 & \frac{R_{\mathrm{f}}}{L_{\mathrm{f}}^2} & 0 & 0 \\
    0 & 0 & 0 & 0 & 0\\
    0 & 0 & 0 & 0 & 0
\end{bmatrix},\
    \bfg{G} = 
    \begin{bmatrix}
        \frac{1}{C_{\mathrm{dc}}} & \frac{-1}{C_{\mathrm{f}}} & \frac{-1}{C_{\mathrm{f}}} \\
        0 & 0 & 0  \\
        0 & 0 & 0 
    \end{bmatrix}.
\end{align}

Thus, dissipative port terms:
\begin{align}
    {\bfg e}_{\bfg R}^{\top} {\bfg f}_{\bfg R}= (\nabla\mathcal{H})^{\top} \left( {\bfg R}\ \nabla\mathcal{H}\right)=R_{\mathrm{f}}\left(\imath_{\mathrm{d}}^2+\imath_{\mathrm{q}}^2\right).
\end{align}

The vector of inputs and outputs influences $\bf y$ is given by:
\begin{align}
    {\bfg y} &= \bfg{G}^{\top} \nabla\mathcal{H} =
    \begin{bmatrix} 
    v_{\mathrm{dc}} & -v_{\mathrm{d}} & -v_{\mathrm{q}} 
    \end{bmatrix}^{\top}.
\end{align}

Thus, port input/output terms are as follows:
\begin{align}
    {\bfg e}_{\bfg{u}}^{\top} {\bfg f}_{\bfg{u}}= {\bfg u}^{\top}  {\bfg y}=\imath^{\star}_{\mathrm{dc}}v_{\mathrm{dc}} -\imath^{\mathrm{g}}_{\mathrm{d}}v_{\mathrm{d}} -\imath_{\mathrm{q}}^{\mathrm{g}}v_{\mathrm{q}}.
\end{align}

The term $\imath^{\star}_{\mathrm{dc}}v_{\mathrm{dc}}$ represents a power source that is controlled through $\imath^{\star}_{\mathrm{dc}}$, and the last two terms denote the power exchanged with the system through the power interface variables: voltage and current. While dissipation is present in the device, it tends to be small relative to the power imbalance.  Therefore {\bf Condition 2 is satisfied} if $\imath^{\star}_{\mathrm{dc}}$ is properly controlled.  This implies that a power reserve must be connected to the dc side.

A Dirac structure can be achieved through the spanned spaces of pairs $(\bfg f_{\bfg{x}},\bfg e_{\bfg{x}} )$, $(\bfg f_{\bfg{u}},\bfg e_{\bfg{u}} )$ and $(\bfg f_{\bfg{R}},\bfg e_{\bfg{R}} )$, ensuring an operation within boundaries and a consistent power interaction between the device and the system. It is worth noting that the matrix $\bfg{J}$ denotes the energy conversion from the dc- to the ac-side and is control-based, as $\bfg{u}^C$ is embedded in $\bfg{J}$. Although the control rules aim to internally balance the stored energy of the dc-side through $v_{\mathrm{dc}}^\star$, they don't adequately address synchronization. While these rules aim to control the voltage magnitude through $v_{\mathrm{ac}}^\star$, they fail to bound its frequency effectively. Therefore, {\bf Condition 3 is not satisfied}.  However, as is shown in Section~\ref{sec_TSC_GFL}, this condition can be satisfied by including frequency control.
\subsection{Grid-forming Converter}
We use the formulation presented in \cite{Baeckeland2025} to model and assess the \ac{gfm} converter, in which the authors present a detailed circuit representation with its respective energy function. 

Minor modifications are applied to the original model and are described next. First, as we do the analysis of the device, we do not take into account the grid. Second, we include the modeling of dc-side similarly to the \ac{gfl} model.

\subsubsection{Model}
For this model we will utilize the equations (\ref{eq_gfl_vdc}) to (\ref{eq_gfl_vq}), which describe the behavior of physical variables such as the dc-bus capacitance, output filter and converter for both, \ac{gfm} and \ac{gfl}. Thus, in terms of physical components, both models are identical. 

Opposite to the \ac{gfl} model, the voltage and current loops are aimed at defining the magnitude and phase of the voltage at the capacitor of the filter, which is located at the terminals of the device and denoted as $v_{\mathrm{d}}=e \, \cos{(\delta)}$ and $v_{\mathrm{q}}=e \, \sin{(\delta)}$. Here, $e$ denotes the voltage magnitude of the voltage at the interconnection terminal, the dq-frame reference is defined by the the state angle $\delta$. Although variables $e$ and $\delta$ are controlled through the Primary Droop, they represent a physical voltage.

As the dq-frame is set by the reference angle $\delta$, the dq-frame frequency $\omega$ can be defined as $\omega= \delta'+\omega_s$.

The dynamic of the controllers is modeled as follows:
\begin{itemize}
    \item  Primary Droop Controller: 
        \begin{align}
            \delta' &= m_p \omega_c \gamma_p, \quad      e^* = e_o + m_q \omega_c \gamma_q, \label{eq_gfm_er}\\
            \gamma_p' &= (p^{\star} - p_h) - \omega_c \gamma_p, \
            \gamma_q' = (q^{\star} - q_h) - \omega_c \gamma_q. \label{eq_gfm_gamma} 
        \end{align}
 
    \item Voltage Controller:
    \begin{align}
        \gamma_{\mathrm{AVC,d}}{'}=&e^{\star}_{\mathrm{d}}-v_{\mathrm{d}}-(1-\rho)K_{\mathrm{w}} \imath^{\star}_{\mathrm{d}}, \\ 
        \gamma_{\mathrm{AVC,q}}{'}=&e^{\star}_{\mathrm{q}}-v_{\mathrm{q}}-(1-\rho)K_{\mathrm{w}} \imath^{\star}_{\mathrm{q}}, \\ 
        \imath^{\star}_{\mathrm{d}}=&\frac{K^{\mathrm{p}}_{\mathrm{AVC}}(e^{\star}_{\mathrm{d}}-v_{\mathrm{d}})+K^{\mathrm{i}}_{\mathrm{AVC}}\gamma_{\mathrm{AVC,d}}}{1-K^{\mathrm{p}}_{\mathrm{AVC}}K^{\mathrm{p}}_{\mathrm{w}}(1-\rho)},\\ 
        \imath^{\star}_{\mathrm{q}}=&\frac{K^{\mathrm{p}}_{\mathrm{AVC}}(e^{\star}_{\mathrm{q}}-v_{\mathrm{q}})+K^{\mathrm{i}}_{\mathrm{AVC}}\gamma_{\mathrm{AVC,q}}}{1-K^{\mathrm{p}}_{\mathrm{AVC}}K^{\mathrm{p}}_{\mathrm{w}}(1-\rho)}.
    \end{align}
    \item Current Controller
    \begin{align}
         \gamma_{\mathrm{CC,d}}{'}=&\rho\, \imath^{\star}_{\mathrm{d}}-\imath_{\mathrm{d}}, \quad
         \gamma_{\mathrm{CC,q}}{'}=\rho\, \imath^{\star}_{\mathrm{q}}-\imath_{\mathrm{q}}, \\
         v^t_{\mathrm{d}}=K_{\mathrm{CC}}^{\mathrm{p}}&(\rho\, \imath^{\star}_{\mathrm{d}}-\imath_{\mathrm{d}})+K_{\mathrm{CC}}^{\mathrm{i}}\gamma_{\mathrm{CC,d}}+v_{\mathrm{d}}- \omega L_{\mathrm{f}} \imath_{\mathrm{q}},\\
         v^t_{\mathrm{q}}=K_{\mathrm{CC}}^{\mathrm{p}}&(\rho\, \imath^{\star}_{\mathrm{q}}-\imath_{\mathrm{q}})+K_{\mathrm{CC}}^{\mathrm{i}}\gamma_{\mathrm{CC,q}}+v_{\mathrm{q}}+ \omega L_{\mathrm{f}} \imath_{\mathrm{d}},\\
         \rho =& \min{\left(1,\frac{\imath_{\max}}{\sqrt{\imath_{\mathrm{d}}^2+\imath^2_{\mathrm{q}}}}\right)}. \label{eq_gfm_rho}
    \end{align}
\end{itemize}

\subsubsection{\ac{tsc} Assessment} \label{sec_Ex:gfm:Assess}

As the components are identical, the storage states defined in (\ref{eq:gfl:x}) and the storage function (\ref{eq:gfl:Hx}) remain valid. Following the same procedure as for the \ac{gfl} case, the control, input and output variables $\bfg u$ can be represented equivalently to the \ac{gfl} case.  

Consequently, the \ac{ph} representation of the \ac{gfm} ensures that the matrices $\bfg{J}$, $\bfg{R}$, and $\bfg{G}$ remain unchanged compared to the \ac{gfl} case. Accordingly, \textbf{Conditions 1 and 2 are satisfied} as long as $\imath^{\star}_{\mathrm{dc}}$ can be properly controlled. 

The key distinction between \ac{gfl} and \ac{gfm} models lies in the definition of the control variables $\bfg{u}^C$, which govern the power transfer from the dc to the ac side. In this scheme, the ac voltage and frequency are directly controlled via a Primary Droop controller. This controller establishes a relation between the active and reactive power at the converter's ac terminals and the frequency and voltage magnitude of the Interconnection Port variables $\bfg{u}^I$, respectively. Thereby, \textbf{Condition 3 is satisfied} if an explicit controller for internal storage balancing is included, as shown in Section~\ref{sec_TSC_GFM}.

\subsection{2nd Order \ac{sm}}
We utilize the 2nd order model of the \ac{sm} where the dynamic of the states $\bfg{x}$ is expressed as:
\begin{align}
    \delta^a\,'= \omega, \qquad
    \omega'= \frac{\tau_{m}-\frac{v_h E}{X_s\omega} \sin(\delta^a-\theta^a_h)}{2H},
\end{align}
where the superscript $a$ denotes absolute angles,
${\bfg x}= 
\begin{bmatrix}
    \delta^a \quad \omega
\end{bmatrix}^{\top}$ and 
$\bfg{u} = 
\begin{bmatrix}
    -E \quad \tau_m
\end{bmatrix}^{\top}.$
In this simplified model we do not model dissipation, thus $ \bfg{R}=\begin{bmatrix} 0_{2 \times 2}\end{bmatrix}$, and the storage function, given by the kinetic energy of the rotor of the \ac{sm}, and its gradient are defined as follows:
\begin{align}
    \mathcal{H}=H\omega^2,\quad  \nabla\mathcal{H}^{\top}=[0\quad 2H\omega].
\end{align}

It is straightforward to conclude that {\bf Condition 1 is satisfied} as the storage function has a quadratic form. 
By factorizing the dynamics by $\nabla\mathcal{H}^{\top}$ we can express the equations as a \ac{ph} system with the following $\bfg{J}$ and $\bfg{G}$ matrices:
\begin{align}
    \bfg{J}=
    \begin{bmatrix}
        0 & \frac{1}{2H}\\
        \frac{-1}{2H} & 0
    \end{bmatrix},  \quad 
    \bfg{G}=
    \begin{bmatrix}
        0 & 0\\
        \frac{v_h \sin(\delta^a-\theta_h)}{X_s\omega 2H} & \frac{-1}{2H}
    \end{bmatrix}.
\end{align}

The vector $\bfg y$ is given by:
\begin{align}
    {\bfg y} &= \bfg{G}^{\top} \nabla\mathcal{H} = \begin{bmatrix}
    \frac{v_h}{X_s}\sin(\delta^a - \theta^a_h) & \omega
    \end{bmatrix}^{\top}.
\end{align}

Thus, port input/output terms are as follows:
\begin{align}
    {\bfg e}_{\bfg{u}}^{\top} {\bfg f}_{\bfg{u}}=
    {\bfg u}^{\top} {\bfg y}=
    -p_h+p_m,
\end{align}
where $p_m = \omega\,\tau_m$ denotes the mechanical input power, and $p_h = \frac{v_hE}{X_s} \sin(\delta^a - \theta^a_h)$ is the active power exchanged with the grid.  If $\bfg{u}^S$ has a constant dynamic, Conditions 2 and 3 will not satisfied, as there is neither dissipation nor internal energy balance control. By introducing damping linked to frequency, {\bf Conditions 2 and 3 are satisfied}, but for a limited range of perturbations, as energy balance and synchronization are inherently supported by the damping.  Nevertheless, in practice, Condition 2 is satisfied when $\bfg{u}^S = \tau_m$ is properly controlled.  Similar to the \ac{gfm} case, Conditions 2 and 3 are closely related, by incorporating a primary droop controller that relates frequency and power, both conditions can be satisfied over a wider range of power disturbances, as the controller actively mitigates frequency drift.

\subsection{Passive Loads}

By definition, a passive inductive or capacitive load has storage capacity related to the magnetic energy stored within the inductance or the electrical energy stored within its capacitance, {\bf satisfying Condition 1}, but in practice its energy tends to be small and might be neglected depending on the time scale. Conversely, a purely resistive load possesses only dissipation capability and no storage. Nevertheless, since none of these load structures possess a controlled input power, they {\bf do not satisfy Condition 2}. Accordingly, it does not provide \ac{tsc}. It is interesting to note that in the case of a constant impedance load, synchronization is achieved without external control or constraints over the voltage and current \cite{Ponce2025}, satisfying Condition 3 partially (as internal energy balance is not achieved inherently) regardless of the power perturbation.

\subsection{Storage Devices}

Consider a system with only batteries as power sources.  While they can provide storage and synchronization, the system will eventually collapse under a sustained power disturbance, as batteries by themselves do not possess \ac{tsc} (Condition 2 is not satisfied).  Nevertheless, it is interesting to consider that if another component in the grid provides a controlled input (i.e., a resource with a large storage capacity and acting in a time scale bigger than that of the batteries), the grid would act as both the interconnection and the controlled input power ports, at different times and probably in different time scales, as the batteries (storage) provide the buffer that bridges the time scales of the power perturbation and the controlled input power.

\section{Case Study}
\label{sec_case}
We utilize a modified version of the WSCC 9-bus system \cite{Sauer_Book}, where the \acp{sm} originally located at bus 1, 2 and 3 have been replaced with \acp{ibr}.
The parameters for all \ac{gfl} and \ac{gfm} models are presented in Table \ref{tab_cases_params}. They remain constant unless explicitly stated otherwise.

\begin{table}[ht]
\caption{\ac{gfl} and \ac{gfm} Controller Parameters}
\centering
\renewcommand{\arraystretch}{1.2}
\begin{tabular}{ll}
\hline
\ac{ibr} Components & Parameters \\
\hline
 dc-capacitor & $C_{\mathrm{dc}}= 0.1$ pu, \\
 Output filter & $L_{\mathrm{f}}=0.0031$ pu , $C_{\mathrm{f}}=0.001$ pu,
  $R_{\mathrm{f}}=0.001$ pu\\
\hline
 \ac{gfl} Controllers & Parameters \\
\hline
 dc-bus voltage & $K_{\mathrm{DVC}}^{\mathrm{p}}=2$, $K_{\mathrm{DVC}}^{\mathrm{i}}=4$, \\
 ac-bus voltage & $K_{\mathrm{AVC}}^{\mathrm{p}}=0.7$, $K_{\mathrm{AVC}}^{\mathrm{i}}=0$, \\
 Current & $K_{\mathrm{CC}}^{\mathrm{p}}=20$, $K_{\mathrm{CC}}^{\mathrm{i}}=200$, \\
 \ac{pll} & $K_{\mathrm{PLL}}^{\mathrm{p}}=10$, $K_{\mathrm{PLL}}^{\mathrm{i}}=100$, \\
\hline
 \ac{gfm} Controllers & Parameters \\
\hline
 Primary Droop & $m_p=0.04$, $m_q=0.04$, $\omega_c=1$~pu, \\
 ac-bus voltage & $K_{\mathrm{w}}^{\mathrm{p}}=0.1$, $K_{\mathrm{AVC}}^{\mathrm{p}}=14.476$, $K_{\mathrm{AVC}}^{\mathrm{i}}=0.273$, \\
 Current & $K_{\mathrm{CC}}^{\mathrm{p}}=9.817$, $K_{\mathrm{CC}}^{\mathrm{i}}=0.018$, $\imath_{\max}=1.5 $~pu. \\
 \hline
\end{tabular}
\label{tab_cases_params}
\end{table}

All simulation results presented in this section were obtained using the simulation software tool Dome \cite{dome_book}.

\subsection{Description of Study-Cases}

\subsubsection{\bf Case 1} 

This is the base case for which we utilise the \ac{gfl} model described by equations (\ref{eq_gfl_vdc}) to (\ref{eq_gfl_omega}). This model is used in the \acp{ibr} located at buses 1, 2 and 3.
\subsubsection{\bf Case 2}
We start from Case 1, but we include a power slack controller to satisfy Condition 2.

Although the control objective regulates the dc-voltage through control laws stated in (\ref{eq_gfl_gammaDVC}) and (\ref{eq_gfl_idref}), it does it by exchanging energy with the system. In this case study, we assume that there is access to energy stored within another time scale, for example a large battery. Thus, we use this energy to regulate voltage at the dc-condenser. Thus, we include a control rule which is dictated by a PI controller, that acts on $\imath^{\star}_{\mathrm{dc}}$ whenever there is a deviation of the dc-voltage from a given reference, as follows:
\begin{align}
    T_{\mathrm{slack}}\imath'^{\star}_{\mathrm{dc}}=&K^{\mathrm{p}}_{\mathrm{slack}}\left(v_{\mathrm{dc}}-v^{\star}_{\mathrm{dc}}\right)+K^{\mathrm{i}}_{\mathrm{slack}}\gamma_{\mathrm{DVC}}-\imath^{\star}_{\mathrm{dc}}, \label{eq_gfl_idcref_case2}
\end{align}
where $T_{\mathrm{slack}}$, $K^{\mathrm{p}}_{\mathrm{slack}}$ and $K^{\mathrm{i}}_{\mathrm{slack}}$ are the time constant and PI gains of the energy controller, respectively. The state of the integral control was defined previously on (\ref{eq_gfl_gammaDVC}). 

\subsubsection{\bf Case 3}
We start from Case 2, but we include a power-frequency controller to satisfy Condition 3.
The current control for the direct axis is now a redundant controller as we assume there is a slack power control for the dc-voltage decribed by (\ref{eq_gfl_idcref_case2}). Thus, we can use this control loop to focus on frequency control. A simplified droop controller is proposed as follows:
\begin{align}
    T_{\omega}\,p^{\star}\,'=K_{\omega}\left(\omega-\omega_s\right)+p-p^{\mathrm{\star}}, \quad
    p^{\star}=v_{\mathrm{d}}\imath^{\star}_{\mathrm{d}}+v_{\mathrm{d}}\imath_{\mathrm{q}}, \label{eq_gfl_idrefomega}
\end{align}
where $T_{\omega}$ and $K_{\omega}$ are the time and droop constant of the frequency control, $p^{\star}$ denotes a power reference, and $\omega$ is an algebraic variable that is defined as the time derivative of the dq-frame angle reference $\theta_e$ in (\ref{eq_gfl_theta}) and (\ref{eq_gfl_omega}). 

Equation (\ref{eq_gfl_idrefomega}) is added and effectively replaces (\ref{eq_gfl_idref}).

\subsubsection{\bf Case 4}
For this case we use the \ac{gfm} model described through the expressions (\ref{eq_gfl_vdc}) to (\ref{eq_gfl_vq}) and (\ref{eq_gfm_er}) to (\ref{eq_gfm_rho}). This model is used in the \acp{ibr} located at buses 1, 2 and 3.

\subsubsection{\bf Case 5}
Starting from Case 4, we include an external dc-voltage controller by introducing a dynamic for the input variable $\imath^{\star}_{\mathrm{dc}}$. To achieve this, we include equation (\ref{eq_gfl_idcref_case2}), enabling the model to satisfy both Conditions 2 and 3.  

\subsubsection{\bf Case 6}
We start from Case 5, but now we consider a primary control based on the \ac{vsm} approach \cite{Zhong2014_Synchronverters} for all the inverters in the grid. Thus, equations (\ref{eq_gfm_er}) and (\ref{eq_gfm_gamma}), are replaced by the following:
\begin{align}
    M\omega' = (p^{\star}-p)-\frac{\omega}{m_p}, \quad \delta' =  \Omega_b(\omega - \omega_s). \label{eq_gfm_swing}
\end{align}

\subsection{\ac{tsc} Analysis for \ac{gfl} Scheme}
\label{sec_TSC_GFL}
\begin{figure}[ht]
    \centering
    \includegraphics[trim=0 10 0 0, clip,width=1.0\linewidth]{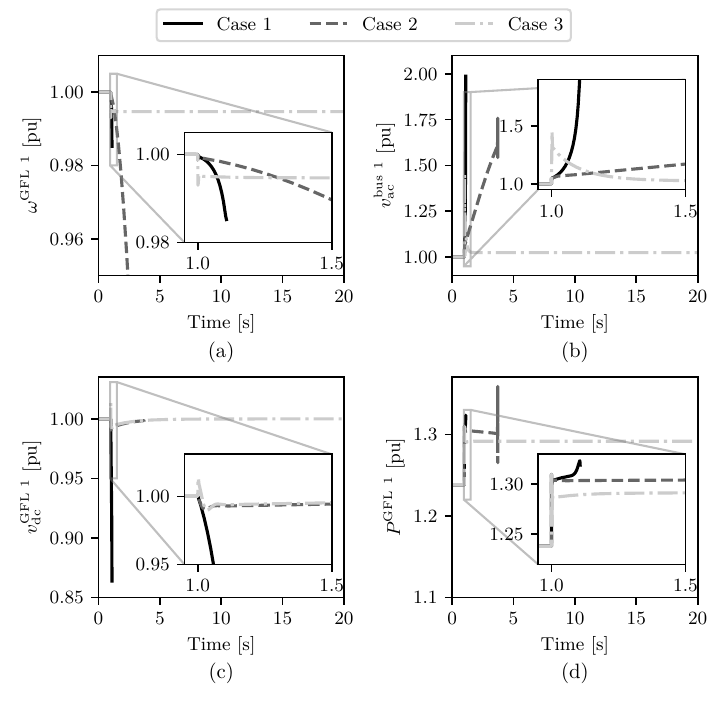}
    \caption{Cases 1, 2 and 3 --- Positive load step at bus 5 ---
    (a) \ac{pll} Frequency $\omega$ for \ac{gfl} at bus 1, (b) ac voltage magnitude at bus 1 $v_{\mathrm{ac}}$, (c) dc voltage for \ac{gfl} 1 $v_{\mathrm{dc}}$ and (d) Active power deliver at bus 1 $p_{\mathrm{1}}$.}
    \label{fig_gfl_wvvp}
\end{figure}

Figure \ref{fig_gfl_wvvp} displays variables of interest for the \ac{ibr} connected to bus 1, following a sustained power perturbation at bus 5. The sustained power perturbation switches on a load equivalent to 5\% of the total load at the first second of the simulation for Cases 1, 2, and 3. Panel (a) displays the frequency estimated by the \ac{pll} at bus 1, panel (b) the magnitude of the voltage at bus 1, panel (c) the dc-capacitor voltage, and panel (d) the output active power delivered at bus 1. All loads are considered constant power, so the contingency is theoretically an ideal power perturbation. A zoom in is included, focused on the first 500 ms transient after the perturbation.
By observing the steady-state values of the dc-bus voltage $v_{\mathrm{dc}}$ in Figure~\ref{fig_gfl_wvvp}.c, as well as the frequency and magnitude of the ac-bus voltage in Figures~\ref{fig_gfl_wvvp}.a and \ref{fig_gfl_wvvp}.b, it can be concluded that Condition~3, i.e., achieving energy balance and synchronization, is satisfied only in Case~3. In Cases~1 and 2, since there is no device providing frequency control, the simulations collapse a few seconds after the perturbation is applied. Therefore, synchronization cannot be achieved in these cases, even though Case~2 is capable of satisfying Condition~2, namely, controlling the dc-bus voltage through a controlled input current.

It is also noteworthy from Figure~\ref{fig_gfl_wvvp}.d that, at the instant the perturbation is applied, power balance is maintained in Cases~1, 2, and 3. This is possible because Condition~1 is satisfied, initially through the energy stored in the filter inductor and the dc capacitor, and subsequently because of Condition~2, access to external energy through the controlled input power.

\subsection{Stored vs. Input Energy}
\label{sec_stored_vs_input}
\begin{figure}[ht]
    \centering
    \includegraphics[trim=0 10 0 20, clip,width=1.0\linewidth]{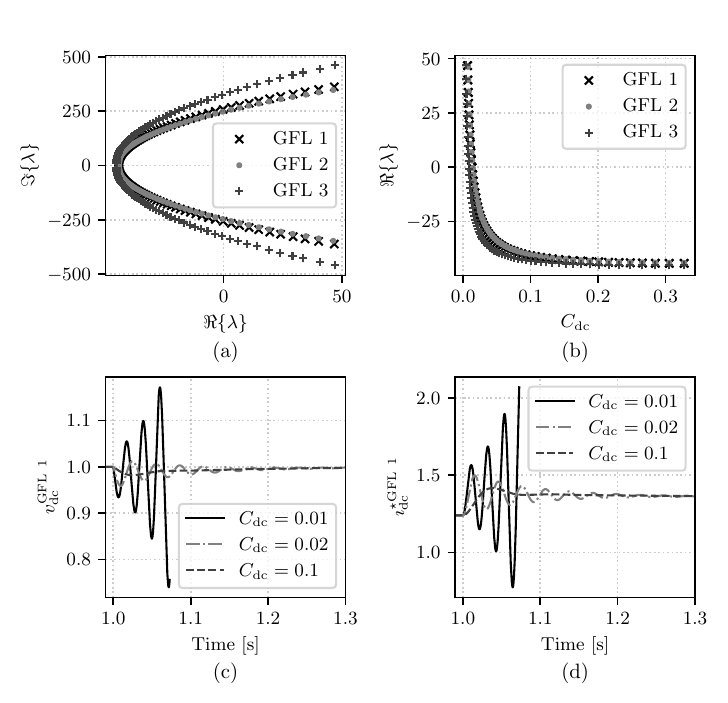}
    \caption{Case 3 --- Positive load step at bus 5 ---
    (a) Real and imaginary parts of critical eigenvalues for different values of $C_{\mathrm{dc}}$ for \acp{ibr} at buses 1, 2, and 3; 
    (b) Real parts of critical eigenvalues as a function of $C_{\mathrm{dc}}$ for \acp{ibr} at buses 1, 2, and 3; 
    (c) dc voltage $v_{\mathrm{dc}}$ for \ac{gfl} 1 for different values of $C_{\mathrm{dc}}$; 
    (d) dc input current $\imath^{\star}_{\mathrm{dc}}$ for \ac{gfl} 1 for different values of $C_{\mathrm{dc}}$.}
    \label{fig_gfl_T01}
\end{figure}

Figure~\ref{fig_gfl_T01}.a illustrates the real and imaginary parts of the critical eigenvalues associated with each \ac{ibr} as the storage capacity of the dc capacitor $C_{\mathrm{dc}}$ varies from 0 to 1~pu. Figure~\ref{fig_gfl_T01}.b complements this analysis by showing only the real parts of the eigenvalues as a function of capacitance. From these figures, it is clear that increasing storage capacity improves small-signal stability by allowing more time for the input power to respond effectively. Conversely, less or no storage capacity leads to instability, as the system lacks the ability to absorb perturbations and provide sufficient buffering for the controlled input power to react.

\begin{figure}[ht]
    \centering
    \includegraphics[trim=0 10 0 20, clip,width=1.0\linewidth]{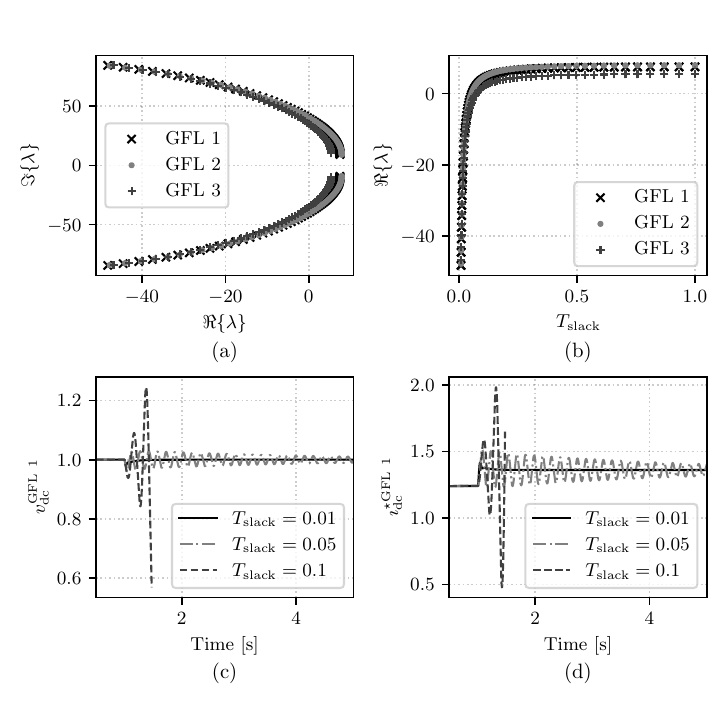}
    \caption{Case 3 --- Positive load step at bus 5 ---
    (a) Real and imaginary part of critical eigenvalues for different values of $T_{\mathrm{slack}}$ for \ac{ibr} at buses 1, 2 and 3, (b) Real part of critical eigenvalues as a function of $C_{\mathrm{dc}}$ for \ac{ibr} at buses 1, 2 and 3, (c) dc voltage for \ac{gfl} 1 $v_{\mathrm{dc}}$ for different values of $T_{\mathrm{slack}}$ and (d)  dc input current for \ac{gfl} 1 $\imath^{\star}_{\mathrm{dc}}$ for different values of $T_{\mathrm{slack}}$.}
    \label{fig_gfl_C1}
\end{figure}

Figures~\ref{fig_gfl_T01}.c and \ref{fig_gfl_T01}.d show the voltage and input current of the dc capacitor evolution in time for three different storage capacity settings, under a power unbalance caused by the connection of a constant power load equivalent to 5\% of the total load. The cases include: $C_{\mathrm{dc}} = 0.01$~pu, which leads to instability; $C_{\mathrm{dc}} = 0.02$~pu, which results in an oscillatory yet stable response; and $C_{\mathrm{dc}} = 0.1$~pu, which leads to a more damped and stable behavior.
A similar assessment is conducted by varying the time constant of the controlled input power, $T_{\mathrm{slack}}$, to evaluate its impact on small-signal stability. Figure~\ref{fig_gfl_C1}.a illustrates the real and imaginary parts of the critical eigenvalues associated with each \ac{ibr} as $T_{\mathrm{slack}}$ varies from 0 to 1~pu. Figure~\ref{fig_gfl_C1}.b complements this analysis by showing only the real parts of the eigenvalues as a function of the time constant. From these figures, it is straightforward to deduce that increasing the controller speed improves small-signal stability by enabling less storage to effectively respond to a sustained power perturbation. Conversely, a slower controller results in instability, as the system lacks the ability to provide a fast enough power slack to reach internal energy balance.

Figures~\ref{fig_gfl_C1}.c and \ref{fig_gfl_C1}.d show the time evolution of the dc voltage and input current for three different time constant settings, under the same 5\% power unbalance. The cases include: $T_{\mathrm{slack}} = 0.1$~pu, which leads to instability; $T_{\mathrm{slack}} = 0.05$~pu, which yields an oscillatory yet stable response; and $T_{\mathrm{slack}} = 0.01$~pu, which results in a more damped and stable behavior.

Comparing Figures~\ref{fig_gfl_T01} and \ref{fig_gfl_C1}, we can observe that there is a dual behavior of the storage capacity with respect to the time constant of the input power. This consistently demonstrates that achieving small-signal stability relies on adequate energy storage and response requirement that can be satisfied in two interchangeable ways: either by increasing the dc‐bus capacitance (Condition 1) to provide longer transient support, or by making the input power controller (Condition 2) to react faster. In practice, controllers can be designed so they can trade off storage size against control speed, larger $C_{\mathrm{dc}}$ relaxes the need for extremely fast power control, while a faster $T_{\mathrm{slack}}$ allows for reduced capacitive storage.

\subsection{\ac{tsc} Analysis for \ac{gfm} Scheme}
\label{sec_TSC_GFM}
\begin{figure}[ht]
    \centering
    \includegraphics[width=0.97\linewidth]{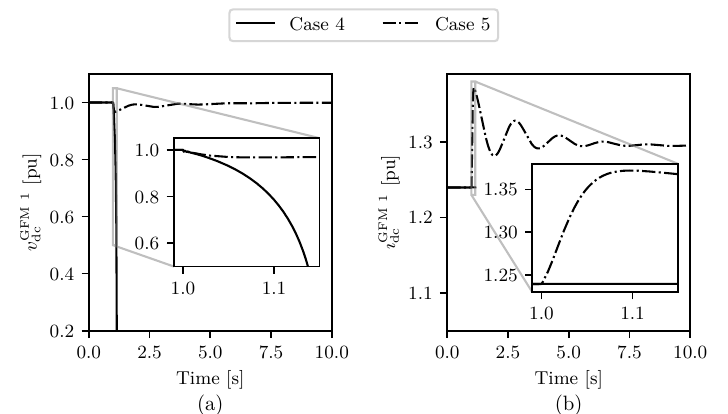}
    \caption{Case 4 and 5 --- Positive load step at bus 5 ---
    (a) dc voltage and (b) dc input current for \ac{ibr} 1.}
    \label{fig_gfm_C4C5}
\end{figure}

In the same vein as the \ac{gfl}, the evaluation of the \ac{tsc} for a \ac{gfm} scheme can be done through the analysis of the dc voltage side of the converter. 
Figure~\ref{fig_gfm_C4C5} shows the voltage at the dc-Capacitor in panel (a) and the input current in panel (b) for Cases 4 and 5. A positive step in the active power load at bus 5, equivalent to 5\% of the total system load, is applied at the first second of the simulation. 
As assessed in Section \ref{sec_Ex:gfm:Assess}, a \ac{gfm} scheme as described in Case 4 cannot provide \ac{tsc} although it possess frequency and voltage magnitude control, since Conditions 2 and 3 are not satisfied. This can be observed in the collapse of the dc-voltage a few milliseconds after applying the perturbation. Nevertheless, by including a controller as the one described in Case 5, internal energy balance can be reached through the input power control, Therefore, Case 5 effectively possess \ac{tsc}.

\subsection{Effect of Limiters in \ac{tsc} for \ac{gfm} Scheme}

A simplified sensitivity can be performed by limiting the output current of the \ac{ibr}. This constraint enforces a decoupling between power and frequency, implying that Condition 3 cannot be fulfilled by the limited \ac{ibr}. This relation is illustrated in Equation~(\ref{eq_gfm_gamma}) for Case 4 and 5, and Equation~(\ref{eq_gfm_swing}) for Case 6.

\begin{figure}[ht]
    \centering
    \includegraphics[trim=0 0 0 35,clip ,width=0.97\linewidth]{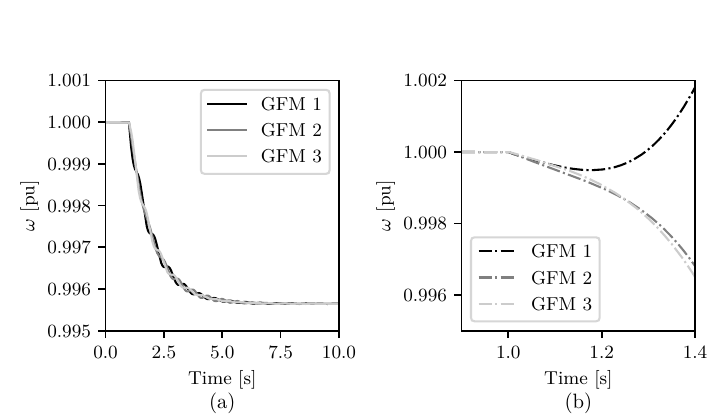}
    \caption{Case 5 --- Positive load step at bus 5 ---
    (a) Frequency of \acp{ibr} 1, 2 and 3 for unlimited current output (b) Frequency of \acp{ibr} 1, 2 and 3 for limited current \ac{gfm} Droop schemes.}
    \label{fig_gfm_C4lim}
\end{figure}

Figure~\ref{fig_gfm_C4lim} shows the time evolution of frequency for the three \acp{ibr} considering Case 5. A positive step in the active power load at bus 5, equivalent to 5\% of the total system load, is applied at the first second of the simulation. Panel (a) represents scenarios with no current limitation, while the panel (b) represents the scenarios in which \ac{ibr} at bus 1 is limited to a maximum current of $\imath_{\max} = 1.415$~p.u.

Although power balance can be maintained by \acp{ibr} located at 2 and 3, the power–frequency control causes the frequency at bus 1 to drift, as the current of the \ac{ibr} located at bus 1, and consequently the active power, becomes limited.

\subsection{Inertia-less \ac{tsc}}

In this subsection, we focus on Case 6, that is, considering a \acp{vsm} scheme for all \acp{ibr}.  Although Condition 1 is satisfied, in practice, the power reserve is provided by the controlled input current as shown in Section~\ref{sec_stored_vs_input}.  The virtual inertia represented by $M$ is in fact a control parameter that links the reference angle $\delta$ with the active power via a second-order dynamic, thereby imitating the physical behavior of a \ac{sm}.

We can reduce the value of $M$ to the point where the system approaches the limiting case in which frequency behaves as an algebraic variable rather than a dynamic state.  Nevertheless, the reference angle $\delta$ always remains well defined. As long as the three conditions are satisfied, the device is capable of driving the system towards a new steady-state energy balance.

\begin{figure}[ht]
    \centering
    \includegraphics[trim=0 5 0 3,clip, width=0.98\linewidth]{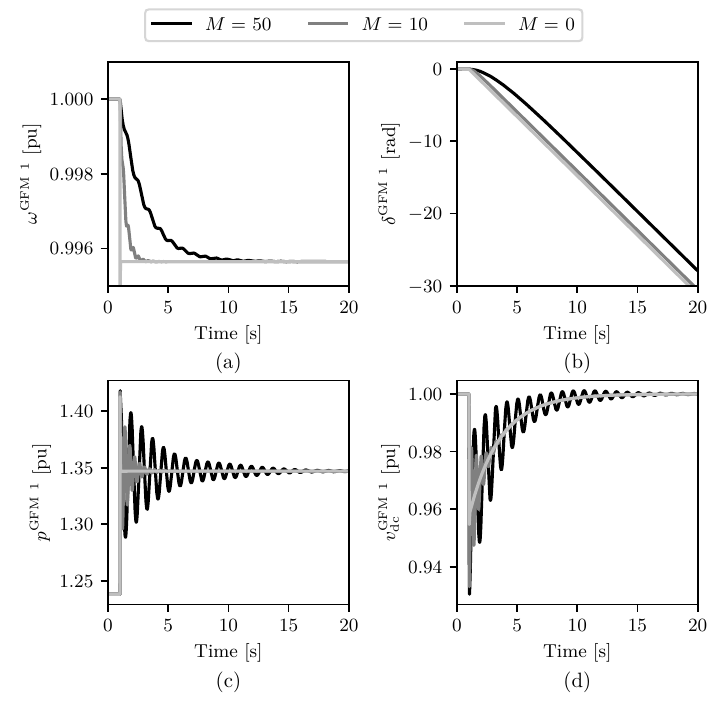}
    \caption{Case 6 --- Positive load step at bus 5 ---
    (a) Frequency, (b) reference angle, (c) active power and (d) dc voltage at bus 1 for \ac{gfm} 1 for different values of inertia $M$.}
    \label{fig_gfm_M}
    \vspace{-3mm}
\end{figure}

To illustrate the impact of virtual inertia on system dynamics, we analyze the system dynamics by applying a 10\% step increase in total load at bus 5, under three different settings of the virtual inertia constant: $M = 50$~s, $M = 10$~s, and $M = 0$~s. Figure~\ref{fig_gfm_M} presents the results: panel (a) shows the frequency at bus 1, panel (b) shows the reference angle $\delta$ at bus 1 (with respect to a 60~Hz rotating frame), panel (c) shows the active power injected by the \ac{ibr} at bus 1, and panel (d) shows the main storage dynamics of \ac{ibr} at bus 1, represented by the voltage across the dc capacitor.
\section{Conclusions and future work}
\label{sec_conclusion}
In this work, we introduce the concept of \ac{tsc}, defined as a set of three necessary conditions based on \ac{ph} formalism that a device, or a set of devices, must satisfy to withstand a sustained power perturbation and drive the system toward a new energy balance.  These conditions are (1) storage capacity; (2) controlled input power; and (3) control driven energy balance.  We also discuss how the relation among the time scales and available energy of the elements that characterize these three conditions is the underlying physical property that defines the proposed \ac{tsc} concept.  Results show that \ac{tsc} is not inherently tied to a specific control scheme, such as, for example, \ac{gfl} and \ac{gfm}.  Future work will focus on the development of new control schemes that satisfy the \ac{tsc} and that go beyond the conventional \ac{gfl}/\ac{gfm} classification.



\end{document}